\documentclass[twocolumn]{aastex62}

\shorttitle{\textit{high-energy retrograde halo stars}}
\shortauthors{Matsuno et al.}

\begin{document}

\title{Origin of the excess of high-energy retrograde stars in the Galactic halo}
\author{Tadafumi Matsuno}
\affiliation{Department of Astronomical Science, School of Physical Sciences, SOKENDAI (The Graduate University for Advanced Studies), Mitaka, Tokyo 181-8588, Japan}
\email{tadafumi.matsuno@nao.ac.jp}
\affiliation{National Astronomical Observatory of Japan (NAOJ), 2-21-1 Osawa, Mitaka, Tokyo 181-8588, Japan}
\author{Wako Aoki}
\affiliation{National Astronomical Observatory of Japan (NAOJ), 2-21-1 Osawa, Mitaka, Tokyo 181-8588, Japan}
\affiliation{Department of Astronomical Science, School of Physical Sciences, SOKENDAI (The Graduate University for Advanced Studies), Mitaka, Tokyo 181-8588, Japan}
\author{Takuma Suda}
\affiliation{Research Center for the Early Universe, The University of Tokyo, 7-3-1 Hongo, Bunkyo-ku, Tokyo 113-0033, Japan}
\begin{abstract} 
We report on the very low $\alpha$-element abundances of a group of metal-poor stars with high orbital energy and with large retrograde motion in the Milky Way halo, whose excess has been reported recently from metallicity and kinematics.
We constructed a sample of halo stars with measured abundances and precise kinematics, including $\sim 880$ stars with [{{Fe}/{H}}]$<-0.7$, by crossmatching the Stellar Abundances for Galactic Archaeology database to the second data release of Gaia. 
Three regions in the energy-angular momentum space have been selected: innermost halo, Gaia Enceladus/Sausage, and high-energy retrograde halo.
While the innermost halo and Gaia Enceladus regions have chemical abundances consistent with high- and low-$\alpha$ populations in the halo, respectively, chemical abundances of stars in the high-energy retrograde halo are different from the two populations; their [{X}/{Fe}], where X represents Na, Mg, and Ca, are even lower than those in Gaia Enceladus. 
These abundances, as well as their low mean metallicity, provide a new support for the idea that the retrograde component is dominated by an accreted dwarf galaxy which has a longer star formation timescale and is less massive than Gaia Enceladus/Sausage.
\end{abstract}


\section{Introduction}
One of the current goals of astronomy is to reconstruct the formation history of the Milky Way.
To this end, signatures of past galaxy accretions are extensively searched for from photometric observations \citep[e.g., ][]{Ibata1994a,Belokurov2006a,Grillmair2006a,Bernard2016} and stellar kinematic information \citep[e.g., ][]{Helmi1999a,Smith2009a,Klement2009a,Helmi2017a,Myeong2018d} in the Galactic halo.
However, it may be difficult to identify a single accretion event from stellar kinematics alone \citep{Jean-Baptiste2017a}.
Therefore, combining chemical and kinematic information is of paramount importance, as the chemical abundances of stars can differ from system to system, for example among dwarf galaxies \citep{Tolstoy2009}. 

There have been suggestions of the existence of two components among Galactic halo stars both in kinematics and chemical abundances \citep[e.g., ][]{Chiba2000a,Carollo2007,Nissen2010}.
Further detailed investigations \added{were} realized thanks to precise measurements of stellar positions, distances, and proper motions by the Gaia mission \citep{GaiaCollaboration2016a}.
A number of studies now draw a consistent picture using Gaia Data Release 1 \citep{GaiaCollaboration2016} and 2 \citep{GaiaCollaboration2018a,Lindegren2018a} that there is a large population of halo stars that show highly eccentric orbits with modest retrograde motion and low-$\alpha$ element abundances, and that they were brought to the Milky Way halo through an accretion of a single massive dwarf galaxy, which is called as Gaia Sausage/Gaia Enceladus \citep[e.g., ][see also an independent result by \cite{Kruijssen2018a}]{Belokurov2018a,Myeong2018a,Koppelman2018a,Deason2018a,Haywood2018a,Helmi2018a}.

A next step is to investigate if we can find other clear accretion signatures.
\citet{Helmi2017a,Myeong2018c} pointed out the excess of stars with high-energy and retrograde orbits using astrometric information from Gaia data \citep{Helmi2017a}, and astrometric and metallicity information from the combination of Gaia and SDSS \citep{Myeong2018c}. 
This excess might be related to a study in the pre-Gaia era, which showed that stars with large retrograde motion have low $\alpha$-element abundances \citep{Venn2004a,Stephens2002}. 
This possible connection should be investigated with the recent astrometric measurements by Gaia and with a large sample of metal-poor stars whose abundances have been measured from high-resolution spectra.

The chemical abundances of $>1,000$ metal-poor stars have been revealed by continuous efforts to identify such stars and measure their stellar abundances.
These abundances are compiled in the Stellar Abundances for Galactic Archaeology (SAGA) database \citep{Suda2008,Suda2011,Yamada2013,Suda2017a}.  

We investigated the current chemo-kinematic view of the stellar halo by combining the SAGA database and Gaia DR2.
In this Letter, we report a new evidence for a past accretion event \citep[e.g., ][]{Venn2004a,Myeong2018c}, confirming its extragalactic origin and strengthening the case that it differs from the ''Gaia Sausage/Enceladus.''
This feature is prominent at low metallicity ([{Fe}/{H}]$\lesssim -1.5$) and has very low $\alpha$-element abundances within the range of $-2.0\lesssim$[{Fe}/{H}]$\lesssim-1.5$, with large retrograde motion.
After describing the sample selection process in Section~\ref{sample}, we present results in Section~\ref{results}. Discussions
 are presented in Section~\ref{discussion}.

\section{Sample\label{sample}} 
\subsection{The SAGA database}
\subsubsection{Chemical abundances}
The abundances of metal-poor stars were extracted from the SAGA database. This database compiles abundances of metal-poor stars from studies that used high- or medium-resolution spectrographs ($R\gtrsim 7,000$).
Given that the density of metal-poor stars on the sky is very low, high-resolution spectroscopic surveys using multi-object spectrographs are not efficient.
Therefore, the use of the SAGA database is an efficient way to obtain chemical abundances of many elements for a large number of metal-poor stars.
We started with $\sim$ 2,100 metal-poor stars ([{Fe}/{H}]$<-0.7$) in this database.

\deleted{In addition to the measurement uncertainties reported in usual abundance studies, there are two major sources of abundance uncertainties in the SAGA database.}
\added{Since our study is based on the abundance data collected from literature, we take two major sources of abundance uncertainties in the SAGA database into consideration.}
One is caused by different methods of abundance analyses among different studies, for example, different stellar parameters or different line lists.
The other is that we mixed various types of stars from main-sequence stars to red giants, between which there could be offsets in abundances caused e.g., by departures from the local thermo-dynamic equilibrium and plane-parallel approximations in real stellar photospheres (non-LTE/3D effects).
Hereafter, we denote $\sigma_1$ and $\sigma_2$ to indicate the contribution from the first and the second effect, respectively.
The total uncertainty $\sigma$ can be expressed as $\sigma^2=\sigma_1^2+\sigma_2^2$.
\added{We note that literature uncertainties are not explicitly adopted in the error estimate here because these uncertainties should be included in the $\sigma$ values evaluated by the following procedure.}

In the following assessments of uncertainties, we used all the stars in the database that have $-3.0<$[{Fe}/{H}]$<-2.5$ and those have $-2<$[{Fe}/{H}]$<-1$.
The $\sigma$ values are expressed as $\sigma_{\rm mp}$ and $\sigma_{\rm mr}$ for the former and the latter sample, respectively.
As seen below, our focus in this \textit{Letter} is the metallicity range $-2<$[{Fe}/{H}]$<-1$, and hence $\sigma_{\rm mr}$ matters.
The $\sigma_1$ was assessed by investigating the median value of standard deviations of abundance measurements for individual objects for which more than two studies had reported abundances.
\added{The $\sigma_{1,\rm mp}$ values (numbers of stars used) are 0.18 $^{+0.06}_{-0.04}$ (20), 0.13 $^{+0.05}_{-0.03}$ (42), 0.08 $^{+0.02}_{-0.03}$ (35), 0.16 $^{+0.07}_{-0.04}$ (43) and 0.10$^{-0.08}_{-0.02}$ (103) for [{Na}/{Fe}], [{Mg}/{Fe}], [{Ca}/{Fe}], [{Ba}/{Fe}], and [{Fe}/{H}], respectively and the $\sigma_{1,\rm mr}$ are 0.07 $^{+0.05}_{-0.02}$ (79), 0.10 $^{+0.03}_{-0.04}$ (97), 0.06 $^{+0.03}_{-0.02}$ (90), 0.18$^{+0.04}_{-0.04}$ (90) and 0.10$^{+0.04}_{-0.04}$ (196).
The superscript and subscript indicate the values between the third quartile and the median and that between the median and the first quartile, respectively.}
We also directly evaluated $\sigma_{\rm mp}$ by examining a spread of [{X}/{Fe}] for each element with the assumption that intrinsic abundance spreads are smaller than measurement errors at $-3<$[{Fe}/{H}]$<-2.5$\footnote{This is not feasible for the metal-rich sample, since abundance ratios are sensitive to the time scale of star formation.}.
\added{We conducted a linear regression and took the half of the difference between 16th and 84th percentiles of residuals as $\sigma_{\rm mp}$.
The $\sigma_{\rm mp}$ values (numbers of stars used) are 0.31 (96), 0.13 (312), and 0.11 (310) for Na, Mg, and Ca\footnote{We use stars with $-2.5<$[{Fe}/{H}]$<-2.0$ to measure $\sigma_{\rm mp}$ for Na since there is a population of extremely metal-poor stars that show very large Na enhancement.}.}
Note that $\sigma_{\rm mp}$ evaluated by this process reflects both two sources of uncertainties.
Thus it is possible to calculate $\sigma_{2,\rm mp}$ from the equation $\sigma_{\rm mp}^2=\sigma_{1,\rm mp}^2+\sigma_{2,\rm mp}^2$ \added{as $\sigma_{2,\rm mp}=0.25,\,0.00,\,0.08$ respectively}\footnote{The above estimate results in $\sigma_{1,\rm mp}$ value comparable to $\sigma_{\rm mp}$ for [{Mg}/{Fe}]. We interpret $\sigma_{2}$ is negligible for [{Mg}/{Fe}] and consider $\sigma_{2,\rm mr}=0$. This would be because of similar ionization potentials of neutral Mg and Fe. }. 
Assuming $\sigma_{2}$ does not depend on metallicity (i.e., $\sigma_{2,\rm mp}=\sigma_{2,\rm mr}$), we get $\sigma_{\rm mr}=$0.27, 0.10, and 0.10 for [{Na}/{Fe}], [{Mg}/{Fe}], and [{Ca}/{Fe}]\footnote{The large metallicity dependence of $\sigma_{1}$ for [{Na}/{Fe}] is probably because Na abundance measurements have to rely on the D lines at low metallicity, which are sensitive to the NLTE effect. $\sigma_{2}$ is also expected to be smaller for high metallicity stars and $\sigma_{\rm mr}$ for [{Na}/{Fe}] is likely to be overestimated.}.
Since there is no way to estimate $\sigma_{\rm mp}$ for [{Fe}/{H}] and thus $\sigma_{2, \rm mp}$ and $\sigma_{2, \rm mr}$, we assumed $\sigma_{\rm mr}=1.5\times\sigma_{1, \rm mr}=0.15$ without estimating $\sigma_2$ values.
\added{It is also not  possible to estimate $\sigma_{\rm mp}$ for [Ba/Fe] due to the intrinsic abundance spread at low metallicity. Therefore, we again skipped the estimation of $\sigma_2$ and assume $\sigma_{\rm mr} = 0.27$ for [Ba/Fe].}
The estimated errors are small enough not to significantly affect our conclusions.

\added{The systematic uncertainties of abundances among different papers are discussed in \citet{Suda2008} where they picked up 17 stars having multiple measurements for carbon abundances, and compared their offsets for the stellar parameters and abundances (in their Fig.10). Possible causes of the uncertainties are also listed, while the inconsistency by the use of different solar abundances from paper to paper is alleviated by the update of the database as discussed in \citet{Suda2017a}.}

\subsubsection{Positions, distances, and proper motions}
Stellar positions and proper motions were obtained from Gaia DR2.
Here, we briefly explain the process of crossmatching the SAGA database to Gaia DR2.
The details of the method will be presented in a forthcoming paper.

We complemented incomplete stellar position data in the database from Simbad using star names
and inspected 2MASS images \citep{Cutri2003} to examine the accuracy of the positions.
After manually correcting the coordinates as required, the SAGA database was crossmatched to 2MASS using the coordinates.
Most of the stars are sufficiently bright to be detected by 2MASS.
Finally, astrometric information was obtained via the \texttt{gaiadr2.tmass\_best\_neighbour} catalog. 
Twenty-five relatively faint stars have no counterparts in the 2MASS point source catalog.
We searched for these 25 objects directly in the \texttt{gaiadr2.gaia\_source} catalog and visually checked the results using Pan-STARRS images. With a few exceptions, the SAGA database was successfully crossmatched with Gaia DR2.
We plan to update the SAGA database to include Gaia information, as well as the kinematics of metal-poor stars.
We adopted the distance estimates of \citet{Bailer-Jones2018a} and further restricted the sample to stars with \texttt{parallax\_over\_error}$>5$.
We also imposed an additional criterion using the equation C.1 of \citet{Lindegren2018a}.
After these processes, 1,571 metal-poor stars remained.

\subsubsection{Radial velocities}
To obtain radial velocities, three sources were combined: Gaia DR2, the SAGA database, and Simbad,
as none of them alone provided radial velocities for a sufficient number of stars.
Radial velocities in the SAGA database and Simbad are based on past measurements in the literature; thus, these sources have heterogeneous data quality. 
The consistency among sources was evaluated by comparing their radial velocity values with those reported in Gaia DR2.
Radial velocity data from the SAGA database were consistent with the measurements obtained by Gaia DR2 at the $2-3\,\mathrm{km\,s^{-1}}$ level; those obtained from Simbad showed similar consistency when using values of quality A or B. 

We established priority in the order of Gaia DR2, SAGA database, and Simbad. We excluded stars that showed significant radial velocity differences ($>10\,\mathrm{km\,s^{-1}}$ corresponding to $\sim 3\sigma$) between different sources; most of them are considered to be in binary systems.
As a result, we were left with 1,290 metal-poor stars that showed no distinct radial velocity variation with good parallax measurements.

\subsubsection{Kinematics}
We used \texttt{galpy} \citep{Bovy2015a} to calculate kinematics of stars. 
We firstly removed disk stars by applying $||\bf{v}-\bf{v}_{\rm LSR}||>180\,\mathrm{km\,s^{-1}}$.
As a result, we have 882 stars, among which 50\% are within $0.88\,\mathrm{kpc}$ and 75\% are within $2.07\,\mathrm{kpc}$.

Energy ($E$) and angular momentum ($L_z$) were calculated adopting a modified \texttt{MWPotential2014} as the Milky Way gravitational potential \citep{Bovy2015a}. 
We replaced the relatively shallow NFW potential in the \texttt{MWPotential2014} with the one with virial mass $M_{200}=1.4\times 10^{12}\,\mathrm{M_\odot}$. 
The concentration parameter was also changed to $c=8.25$ to match the rotation curve of Milky Way (private comm. with K. Hattori).
We subtracted the potential energy at a very large distance from the obtained $E$ to get $E=0$ at an infinite distance from the Galactic center, as explained in the document of \texttt{galpy}. 
The obtained $E-L_z$ distribution is presented in Figure \ref{figkine}.

\subsection{LAMOST DR4}
Since the number of stars in the database is still not very large and since there is a clear bias toward metal-poor stars in the database, we also investigate the $E-L_z$ distribution of metal-poor A-, F-, G-, and K-type stars catalogued in LAMOST DR4 \citep[][lower panel of Figure~\ref{figkine}]{Cui2012a}.
We simply crossmatched stars in LAMOST estimated to be [{Fe}/{H}]$<-0.7$ to Gaia DR2, and selected halo stars with the same criteria as those used for the SAGA database stars. 
We have 35069 stars from LAMOST, and 50\% of stars are within $1.75\,\mathrm{kpc}$ and 75\% are within $2.63\,\mathrm{kpc}$.
We just used these LAMOST stars to confirm the $E-L_z$ distribution of stars in the SAGA database and to investigate metallicity distributions of selected regions.

\subsection{Selection boxes}
\begin{figure*}
\plotone{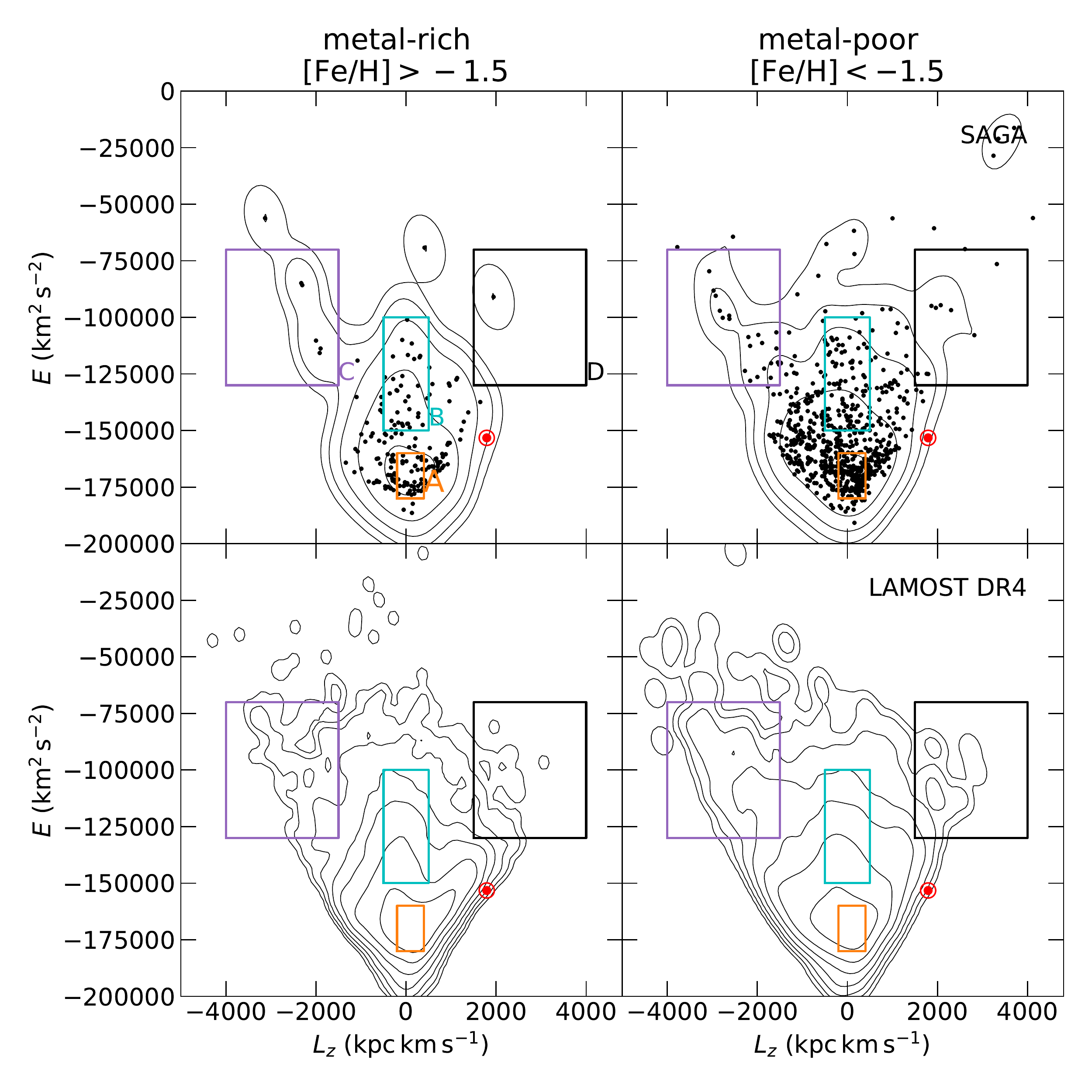}
\caption{Distribution of stars in the energy ($E$) -- angular momentum ($L_z$) space after dividing by the metallicity [{Fe}/{H}]$=-1.5$, for stars in the SAGA database (\textit{upper} panel) and in A-, F-, G-, and K-type stars catalogued in LAMOST DR4 (\textit{lower} panel). Individual stars in the SAGA database are plotted, as well as the contour; for LAMOST stars, only the contour is shown. The rectangles show the four regions used in subsequent chemical analyses (Table~\ref{tabkinediv}). The location of the Sun is also shown by red circles. \label{figkine}}
\end{figure*}

\begin{figure}
\plotone{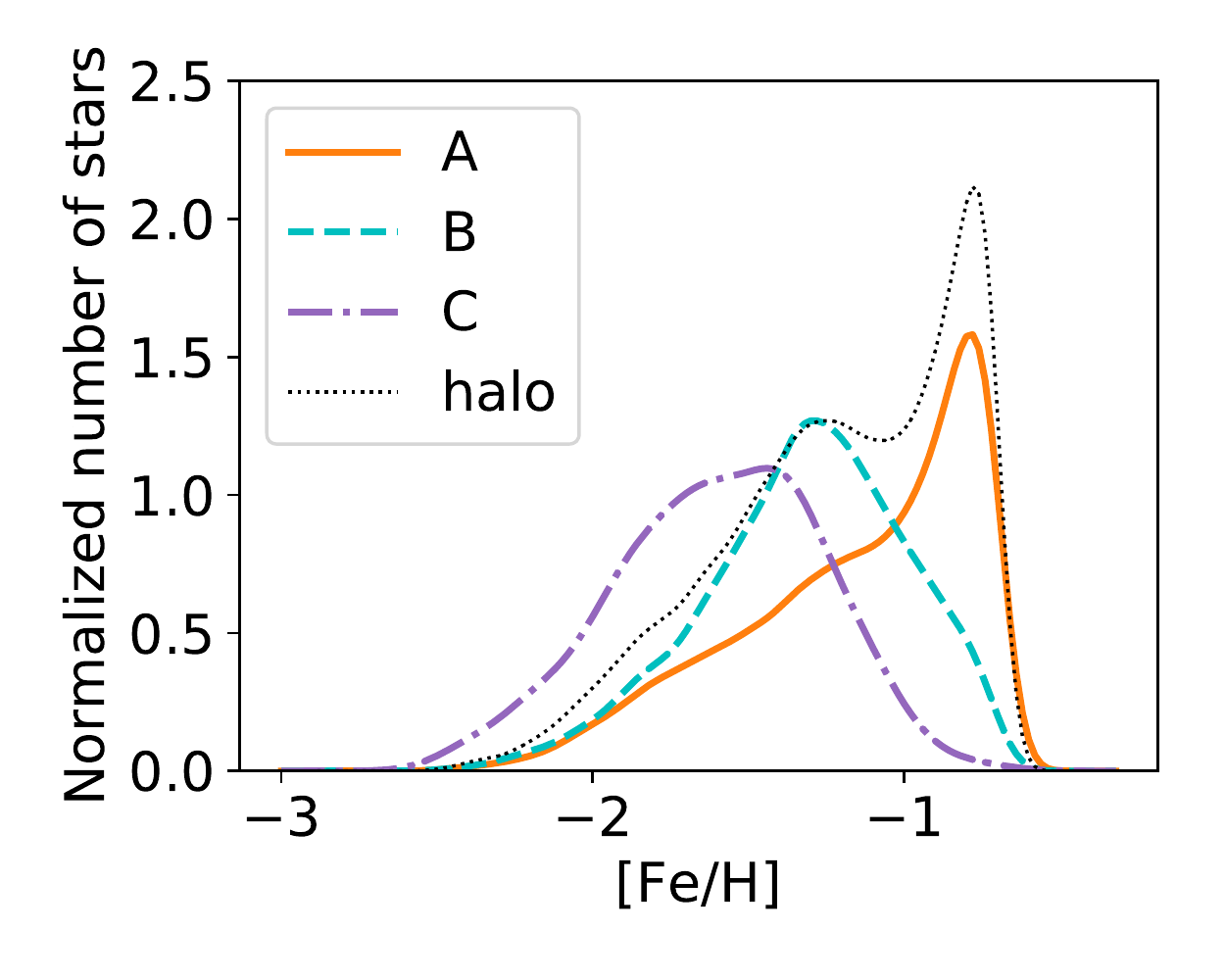}
\caption{Normalized metallicity distribution functions of stars in the three selected regions and that of all the halo stars. This figure was constructed with Gaussian kernel density estimator using the A, F, G and K stars catalogued in LAMOST DR4, without the SAGA database. Note that the sharp cut-offs at high metallicity are due to our sample selection with [{Fe}/{H}]$<-0.7$ and that the histogram for the entire halo is multiplied by 1.5 for the visualization purpose. \label{figmdf}}
\end{figure}

\begin{deluxetable*}{llrrrrrrr}
\tablecolumns{9}
  \tablecaption{Properties of the four selected regions\label{tabkinediv}}
  \tablehead{
\colhead{Region ID} & \colhead{Name} & \multicolumn{2}{c}{\# of stars}  &\multicolumn{2}{c}{$\langle L_z\rangle\pm\sigma_{L_z}$}                   &\multicolumn{2}{c}{$\langle E\rangle\pm\sigma_{E}$}                   & \colhead{$\langle$[{Fe}/{H}]$\rangle\pm\sigma_{\rm [{Fe}/{H}]}$} \\
                    &                & \colhead{SAGA}&\colhead{LAMOST}  &\colhead{SAGA}&\colhead{LAMOST}                                           &\colhead{SAGA}&\colhead{LAMOST}                                           &  LAMOST    \\
                    &                &          &                       &\dcolhead{\times 10^3\mathrm{kpc\,km\,s^{-1}}} &\dcolhead{\times 10^3 \mathrm{kpc\,km\,s^{-1}}}  &\dcolhead{\times 10^5\mathrm{km^2\,s^{-2}}} &\dcolhead{\times 10^5\mathrm{km^2\,s^{-2}}}  &
}
\startdata
A                   & Innermost halo         &   161                   & 8954                   & $0.10 \pm 0.17$   &$0.12 \pm 0.17$                 & $-1.71\pm 0.05$ &$-1.70\pm 0.05$   & $-1.16 \pm 0.38$ \\
B                   & Gaia Enceladus         &   135                   & 4222                   & $-0.02\pm 0.27$   &$-0.06\pm 0.26$                 & $-1.34\pm 0.13$ &$-1.36\pm 0.11$   & $-1.32 \pm 0.33$ \\
C                   & high-$E$ retrograde    &   26                    & 299                    & $-2.17\pm 0.48$   &$-2.20\pm 0.52$                 & $-1.08\pm 0.14$ &$-1.06\pm 0.15$   & $-1.60 \pm 0.33$ \\
D                   & \nodata                &   10                    & 70                     & $ 2.14\pm 0.54$   &$1.93 \pm 0.35$                 & $-1.03\pm 0.17$ &$-1.14\pm 0.12$   & $-1.46 \pm 0.47$ \\
\enddata
\end{deluxetable*}

In Figure~\ref{figkine}, we show the distribution of stars with [{Fe}/{H}]$<-0.7$ in the SAGA database in the $E$--$L_z$ plane. 
The contour was made using a Gaussian kernel density estimator.
The upper panels show that the stellar kinematic properties vary with metallicity.
At higher metallicity ([{Fe}/{H}]$>-1.5$; upper left panel), 
we see the signature of Gaia Enceladus/Sausage at $L_z\sim -500\,\mathrm{kpc\,km\,s^{-1}}$ and $E>-1.6\times 10^5\,\mathrm{km^2\,s^{-2}}$ \citep{Belokurov2018a,Myeong2018a,Koppelman2018a,Deason2018a,Haywood2018a,Helmi2018a}.
Gaia Enceladus is interpreted as the result of dwarf galaxy 
accretion.  
As we move toward lower metallicity ([{Fe}/{H}]$<-1.5$), the Gaia Enceladus signature becomes 
weak \citep{Belokurov2018a,Myeong2018c}.
Instead, we see a clear enhancement of stars with retrograde motion. 
This metallicity difference between Gaia Enceladus and high-energy retrograde halo stars seems consistent with Figure 2 of \citet{Myeong2018c}, who noted that the excess of high-energy retrograde stars extends down to [{Fe}/{H}]$\sim -1.9$ while the diamond shape in the $L_z-E$ space, corresponding to Gaia Sausage/Enceladus, extends down to [{Fe}/{H}]$\sim-1.5$.

The star distributions in LAMOST DR4 are similar to those in the SAGA database; the basic picture described above was confirmed by the LAMOST DR4 sample. 
Slight differences are attributable to the small number of stars in the SAGA database, the different metallicity distributions between the two samples, and/or the radial velocity and metallicity measurement quality. The SAGA database focuses on lower metallicity and has smaller uncertainties in radial velocity and metallicity measurements.

In the following chemical analysis, we compare the abundances of stars in the four regions in the $E$--$L_z$ plane, shown by the rectangles in Figure~\ref{figkine} (see also Table~\ref{tabkinediv}). 
The first three regions in $E$--$L_z$ are the innermost halo with small $E$ and prograde motion (orange labeled as A), Gaia Enceladus with \deleted{large $E$ and slightly retrograde motion}\added{high $E$ and low $L_z$} (cyan; B), and the high-energy retrograde stars (purple; C). 
The selection box C roughly corresponds to S1, Rg2, Rg3, Rg4, and Rg6 of \citet{Myeong2018b}. \footnote{Since our analysis and that of \citet{Myeong2018b} are different, the comparison is not very precise. However, we note that we obtain similar $L_z$ and $E$ for $\omega$ Centauri $(-595\,\mathrm{kpc\,km\,s^{-1}},-1.78\times10^5\,\mathrm{km^2\,s^{-2}})$ to their values $(-496\,\mathrm{kpc\,km\,s^{-1}},-1.85\times10^5\,\mathrm{km^2\,s^{-2}})$. }
The last region, with a high $E$ and prograde motion, was selected for the region C comparison (black; D).
We note that results presented below are unchanged if we change the boundary $L_z$ by a few $\times 100\,\mathrm{kpc\,km\,s^{-1}}$ or $E$ by $\sim 10^4\,\mathrm{km^2\,s^{-2}}$ of the selection boxes.

\section{Results\label{results}}
\begin{figure*}
\plotone{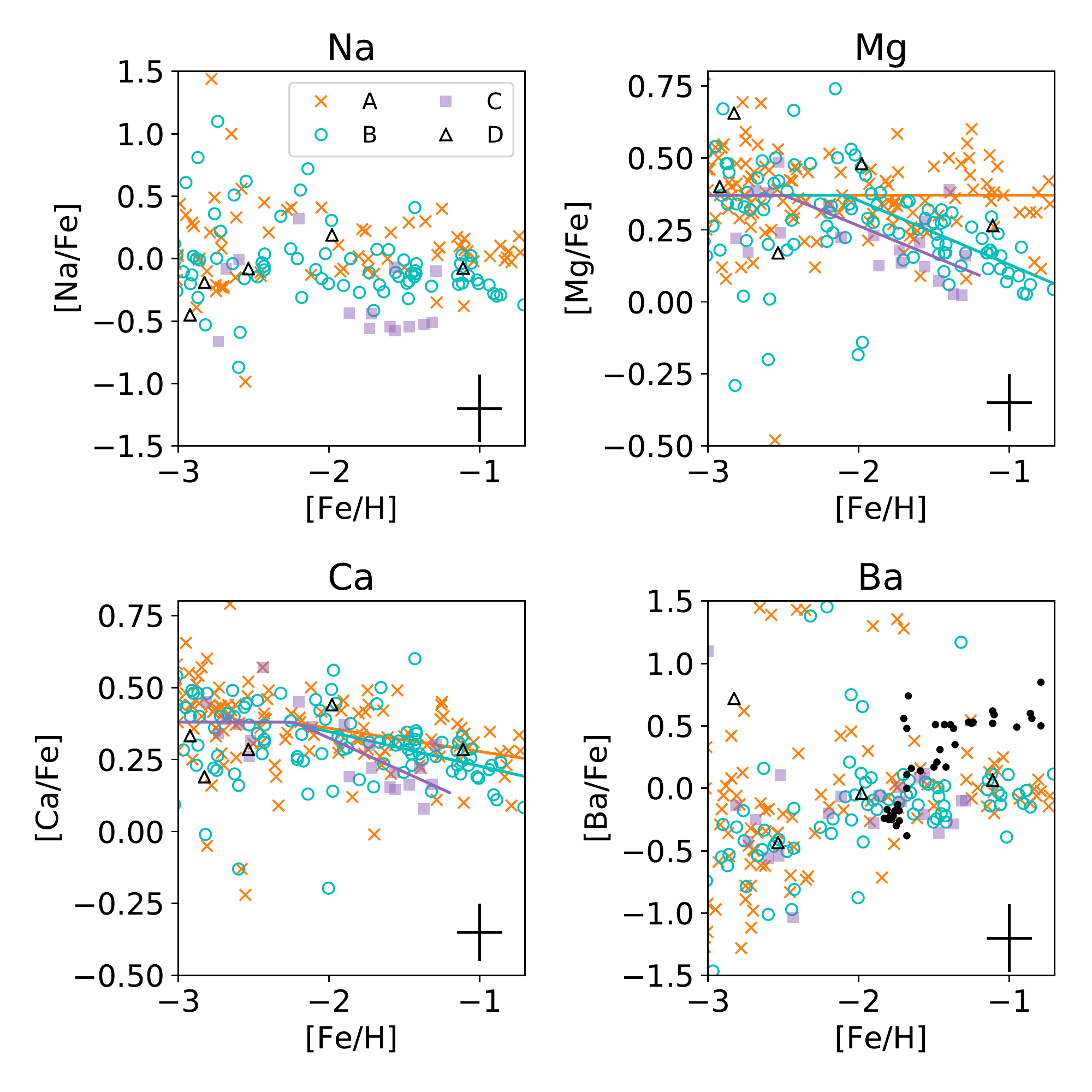}
\caption{Chemical abundances of stars in the four regions for Na, Mg, Ca, and Ba. The data are taken from the SAGA database. Small black dots in the [{Ba}/{Fe}] panel are stars in $\omega$ Cen from \citet{Norris1995}. Regions A--C appear to occupy different positions in each of the three panels for Na, Mg and Ca. Note that the vertical scales for Na and Ba are different from the others. The lines in Mg and Ca panels show approximate chemical evolution of regions A--C. See Section \ref{discussion} and Figure \ref{figabun2} for more details. \label{figabun1}}
\end{figure*}

Figure~\ref{figmdf} shows metallicity distributions of stars in the three regions from the LAMOST DR4 catalog. It is very clear that the three regions (A-C) have different metallicity distributions.
The innermost halo (A) has the highest metallicity, while the retrograde substructure (C) has the lowest. 
In addition to this metallicity difference, we investigated abundance trends in detail in the following.

Figure~\ref{figabun1} shows the chemical abundance trends of stars in the four regions for Na, Mg, Ca and Ba from the SAGA database; notably,
data points that had only upper limits were excluded.
This did not affect Na, Mg and Ca at [{Fe}/{H}]$>-3.0$ and only one star belonging to the innermost region was excluded ([{Fe}/{H}]$=-2.56$ and [{Ba}/{Fe}]$<-1.32$). 
When a star had multiple measurements for a given element, we simply took the average of the values for plotting.

It is known that there are two distinct chemical populations in the Galactic halo, namely high-/low-$\alpha$ populations \citep[e.g., ][]{Nissen2010}. 
\citet{Nissen2010,Nissen2011} showed that the high-$\alpha$ population has higher [{X}/{Fe}] for the three elements, Na, Mg, and Ca.
Recent analyses of halo stars successfully associated the low-$\alpha$ population with the Gaia Enceladus from kinematics and chemical abundances of stars \citep[e.g., ][]{Haywood2018a,Helmi2018a}.
This chemical abundance difference is understood as a result from slower star formation in the low-$\alpha$ population. 
This slower star formation leads to lower metallicity by the time of onset of type Ia supernovae.

Figure~\ref{figabun1} confirms lower-$\alpha$ abundances of Gaia Enceladus (B) relative to the innermost halo population (A). 
A striking feature shown in the figure is that the retrograde substructure (C) does not follow either of the overall abundance trend of Gaia Enceladus or that of the innermost halo, with even lower [{X}/{Fe}] of the three elements on average than those of Gaia Enceladus at [{Fe}/{H}]$\gtrsim -2.0$.
This indicates that the retrograde halo has a progenitor that is independent of the innermost halo or Gaia Enceladus. 
We further discuss the properties of the high-energy retrograde halo stars in the next section from the perspective of chemical abundance.

Region D was selected for the comparison.
It has the same range of $E$ as high-energy retrograde halo stars, but with prograde motion.
Therefore, the region D provides us with estimates of the contribution of the ``smooth'' component of the halo to the region C.
The region D does not have many stars at [{Fe}/{H}]$>-2.5$ as the region C, and a few stars with [{Fe}/{H}]$>-2.5$ have different abundances from most of the stars in the region C.
This indicates that the high-energy retrograde halo stars clusters in both kinematic \citep{Myeong2018b,Myeong2018c} and chemical space and represents a distinct population.

For completeness, note that although we investigated other elements (C, Ti, Zn, Sr, Y, Ba, and Eu), we did not see significant differences among the regions, with Zn being an exception such that it might show a hint of possible abundance difference between high-energy retrograde stars and Gaia Enceladus. Although the lacks of the difference may be partially due to insufficient precision of measured abundances, intrinsic abundance scatter of neutron capture 
elements, and/or abundance change during the stellar evolution, Ba anomaly such as seen in $\omega$ Centauri \citep{Norris1995} is clearly absent among the high-energy retrograde stars (lower right panel of Figure~\ref{figabun1}).

\section{Discussion \label{discussion}}
\begin{figure*}
\plotone{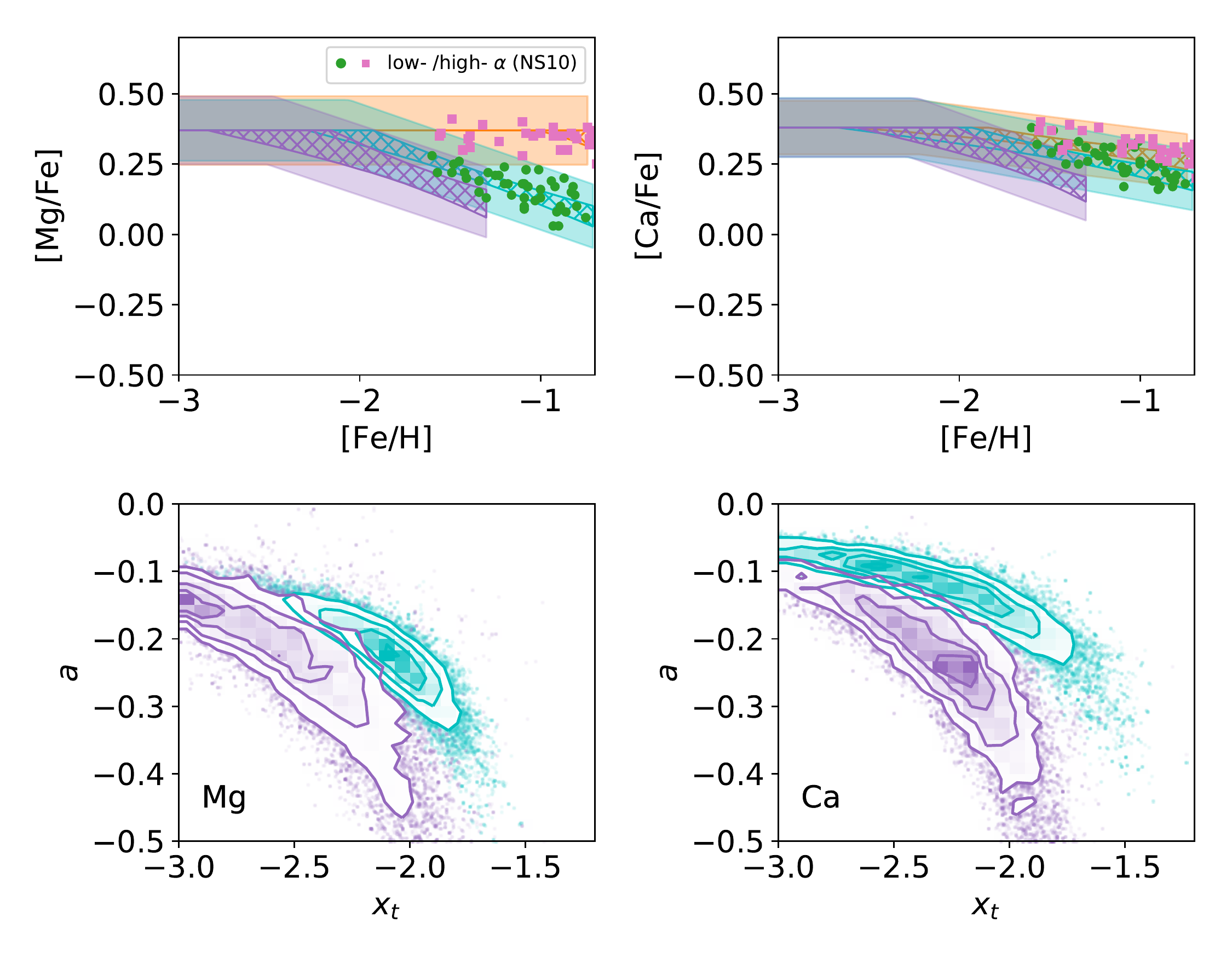}
\caption{\textit{upper panel}: Chemical abundance trends of stars in the three regions in the Milky Way halo for Mg and Ca. Colors for the three regions follow those in Figure~\ref{figabun1}. Halo stars from \citet[][; green/salmon symbols corresponding to low-/high-$\alpha$]{Nissen2010,Nissen2011} are plotted for comparison purposes. \textit{lower panel}: Posterior distributions of the obtained parameters for Gaia Enceladus and high-energy retrograde halo stars. $x_t$ denotes the metallicity at which [{Mg}/{Fe}] or [{Ca}/{Fe}] starts decreasing, and $a$ denotes the slope of the trend at [{Fe}/{H}]$>x_t$ (see text and equations \ref{eq1} and \ref{eq2}). \label{figabun2}}
\end{figure*}

We approximated Mg and Ca abundance trends with the following form of function for chemical evolution (Figures \ref{figabun1} and \ref{figabun2}),
\begin{equation}
f(x)=
  \left\{
    \begin{array}{ll}
y_0 & (x<x_t) \\
a (x-x_t) + y_0 & (x>x_t) 
    \end{array}
  \right.\label{eq1} 
\end{equation}
where $x,y$ are for [{Fe}/{H}] and [{X}/{Fe}]. 
To obtain the set of parameters which describes the data best, we adopted the following likelihood, 
\begin{equation}
p = \Pi_i \int f(\xi|\mathrm{[Fe/H]}_i,\sigma_{\rm [Fe/H]})g(\mathrm{[X/Fe]}_i|\xi,\bf{x})d\xi
\end{equation}
where 
\begin{eqnarray}
f =&\frac{1}{\sqrt{2\pi\sigma^2_{\rm [Fe/H]}}}\exp (-\frac{(\xi - \mathrm{[Fe/H]}_i)^2}{2\sigma_{\rm [Fe/H]}^2})\\
g =& (1-f_o)\frac{1}{\sqrt{2\pi \sigma_{\rm [X/Fe]}^2}}\exp(-\frac{([\mathrm{X/Fe}]_i - f(\xi))^2}{2\sigma_{\rm [X/Fe]}^2}) \nonumber\\ &+  f_o\frac{1}{\sqrt{2\pi (\sigma_{\rm [X/Fe]}^2+\sigma_b^2)}}\exp (-\frac{([\mathrm{X/Fe}]_i-\mu_b)^2}{2(\sigma_{\rm [X/Fe]}^2 + \sigma_b^2)}).\label{eq2} 
\end{eqnarray}
$f_o,\,\mu_b$, and $\sigma_b$ are outlier fraction, mean and standard deviation for the outlying population. 
\deleted{We estimated a set of parameters $(a,x_t,f_o,\mu_b,\sigma_b)$ using MCMC sampling while fixing $(y_0,\sigma_X)$.}
\added{We estimated a set of parameters $(a,x_t,\sigma_{\rm [X/Fe]},f_o,\mu_b,\sigma_b)$ using MCMC sampling while fixing $y_0$ and $\sigma_{\rm [Fe/H]}$.}
\deleted{Errors estimated in the previous section were adopted as $\sigma_X$ ($0.10$ for Mg and $0.07$ for Ca) and }The mean [{X}/{Fe}] in $-3.0<$[{Fe}/{H}]$<-2.5$ were adopted as $y_0$ ($0.37$ for Mg and $0.38$ for Ca) \added{and $\sigma_{\rm [Fe/H]}=0.15$ was adopted}.
\deleted{Flat prior with a sufficiently wide range were adopted for all the parameters but $f_o$, for which flat prior between 0 and 0.5 was used.}
\added{Flat priors with sufficiently wide ranges were adopted for the parameters except for $x_t$ ($-3<x_t<-1$) and $f_0$ ($0<f_0<0.5$).}

Posterior distributions for $a$ and $x_t$ are shown in the lower panels of Figure~\ref{figabun2}.
The posterior distributions show that Gaia Enceladus and high-energy retrograde stars are fit with different sets of parameters.
\deleted{We note that if we allow the $\sigma_X$ to change, derived $\sigma_X$ result in the value comparable to the estimated error, indicating each region forms a tight chemical sequence.}
\added{The $\sigma_X$ result in comparable to the estimated errors ($\sigma_{\rm[Mg/Fe]}=0.12,\,0.11,\,\mathrm{and}\,0.12$ and $\sigma_{\rm[Ca/Fe]}=0.10,\,0.11,\,\mathrm{and}\,0.10$ for regions A, B, and C respectively), indicating abundance spread of each region is smaller than or comparable to the estimated errors.}
\added{We note that $f_0$ converge between $0.10-0.20$ for the regions A and B, and $<0.10$ for the region C.}

The best models are shown in Figure \ref{figabun1} and the upper panels of \ref{figabun2}. The widths of the shaded areas correspond to \deleted{the errors previously estimated for each point ($\sigma_X$ in equation \ref{eq2})}\added{$\sigma_{\rm [X/Fe]}$ and $\sigma_{\rm [Fe/H]}$ and the hatched areas correspond to $1\sigma$ regions of the best fit model (equation \ref{eq1}).}
Also shown \added{in Figure \ref{figabun2}} are ``low-$\alpha$'' and ``high-$\alpha$'' populations of \citet{Nissen2010,Nissen2011}.
Note that we did not include region D in the following analysis, as it is not associated with main features in the $E$--$L_z$ plane.

Figure~\ref{figabun2} confirms that our innermost halo and Gaia Enceladus stars correspond to the high-/low-$\alpha$ populations of \citet{Nissen2010}, respectively.
The general interpretation of the two populations is that the high-$\alpha$ population experienced more intense star formation prior to the onset of type~Ia supernovae. 
To achieve such a high star formation rate at the early phase, the high-$\alpha$ population is usually considered to have formed in a massive galaxy, probably the Milky Way itself, although the detailed process is still under debate \citep[e.g., ][]{Fernandez-Alvar2018a,Mackereth2018a}.
On the other hand, the low-$\alpha$ population of \citet{Nissen2010} is now considered to be an accreted dwarf galaxy (Gaia Enceladus) from chemical abundances and kinematics \citep[e.g., ][]{Helmi2018a,Belokurov2018a,Haywood2018a}. 

Figure~\ref{figabun2} also shows that the [{X}/{Fe}] ratios in high-energy retrograde halo stars are even lower for the \deleted{three}\added{two} elements than the two halo populations in \citet{Nissen2010} at [{Fe}/{H}]$\gtrsim -2$. 
High-energy retrograde halo stars has been enriched only up to [{Fe}/{H}]$\sim -2.5$ by the time of onset of type~Ia supernovae, which indicates slow star formation.

The slow star formation indicated from the very low-$\alpha$ element abundances suggests very inefficient star formation, which would suggest a low mass progenitor.
The mass ratio between the progenitor of high-energy retrograde halo stars and Gaia Enceladus was estimated by their metallicity distribution functions (Figure~\ref{figmdf}).
The mean metallicity of Gaia Enceladus is $\sim -1.3$, and that of high-energy retrograde halo stars is $\sim -1.6$.
The mass-metallicity relation of \citet{Kirby2013} for dwarf galaxies suggests that this $0.3\,\mathrm{dex}$ difference corresponds to a factor of $\sim 10$ stellar mass difference. 

Considering this large mass ratio, the impact of the accretion of the progenitor of high-energy retrograde halo stars to Milky Way is likely to be much smaller than that of Gaia Enceladus. 
However, such a small system is still detectable by kinematics \citep{Myeong2018c} and addition of chemical abundance information brings us robust conclusion and tells us the property of the progenitor.
Note that we did not find many stars in the SAGA database that are similar in chemical abundances, but not in kinematics, to stars in high-energy retrograde halo stars.
Therefore, high-energy retrograde halo stars would be a unique contributor to the Milky Way stellar halo.

\citet{Myeong2018b} discussed a possible connection of some of their high-energy retrograde substructures with $\omega$ Centauri.
The abundance pattern of high-energy retrograde halo stars is different from the stellar chemical abundances in the globular cluster reported by \citet{Johnson2010a}; they reported almost flat $\alpha$-element abundances up to [{Fe}/{H}]$\sim -1$ for $\omega$ Centauri.
This difference as well as the lack of Ba abundance anomalies \added{(lower right panel of Figure~\ref{figabun1})} indicates that the majority of high-energy retrograde halo stars is unrelated to $\omega$ Centauri.

Considering that we only used a compilation of past abundance measurements, which can be affected by systematic uncertainties, sufficient precision could be achieved in large spectroscopic surveys with a well-calibrated analysis if the surveys are designed well to study metal-poor stars.
Indeed, we have reached a consistent conclusion for Mg using APOGEE DR14 data \citep{Holtzman2015}.
However, other elements in APOGEE do not show as clear differences as Mg. 
This is due to the limitations of the current surveys in terms of the number of halo stars and the accuracy of chemical abundance measurements for metal-poor stars; additionally, it 
highlights the need for high-resolution spectroscopic surveys designed specifically to study halo stars.

\section{Summary}

Based on chemical abundances and kinematics from the SAGA database and Gaia DR2, we added new evidence that the excess of stars with highly-retrograde orbits at high energy is caused by an accretion of a dwarf galaxy which is different from Gaia Enceladus/Sausage. 
Compared to previous studies that have pointed out or investigated the excess with stellar kinematics and metallicity \citep{Helmi2017a,Myeong2018c,Myeong2018b}, we included $\alpha$-element abundances in the investigation. 
The $\alpha$-element abundances are even lower than the low $\alpha$-element abundances of Gaia Enceladus, suggesting a different and lower mass progenitor.
Although there are studies that pointed out stars with large retrograde motion have low $\alpha$-element abundances \citep{Venn2004a,Stephens2002}, these studies were in the pre-Gaia era, and hence our study is new in that it used the latest most precise kinematics from Gaia astrometric measurements and discussed the population in connection with a recently identified accretion signature in the Galaxy.

Our results are based on simple kinematic division.
For a more detailed discussion, it is necessary to construct a pure sample for each component based on the abundance ratios. 
It would also be interesting to investigate the abundances of many elements for high-energy retrograde halo stars with high precision using high-quality spectra, for example, to include neutron capture elements and compare them with stars in surviving dwarf galaxies around the Milky Way.
These approaches are not yet feasible, due to the limited number of stars and the limited precision of abundances, as we used a compilation of past abundance measurements from the literature. 
We plan to carry out high-precision abundance analyses similar to those of \citet{Nissen2010} in the future.

\acknowledgements
We thank the referee for the detailed review of the paper.
We thank Timothy C. Beers and Kohei Hattori for their helpful suggestions and comments.
TM is supported by Grant-in-Aid for JSPS Fellows (Grant number 18J11326).
WA and TS were supported by JSPS KAKENHI Grant Numbers 16H02168, 16K05287, and 15HP7004.
This work has made use of data from the European Space Agency (ESA)
mission {\it Gaia} (\url{https://www.cosmos.esa.int/gaia}), processed by
the {\it Gaia} Data Processing and Analysis Consortium (DPAC,
\url{https://www.cosmos.esa.int/web/gaia/dpac/consortium}). Funding
for the DPAC has been provided by national institutions, in particular
the institutions participating in the {\it Gaia} Multilateral Agreement.

\end{document}